\documentclass[aps,pra,twocolumn,showpacs,amsmath,amsmath,amssymb]{revtex4-1}

\usepackage{amsmath,amssymb,graphicx,bm,psfrag,bbold,float,color}

\usepackage[mathscr]{eucal}

\DeclareMathAlphabet{\mathpzc}{OT1}{pzc}{m}{it}

\begin{document}

\title{Charge Qubit-Atom Hybrid}

\author{Deshui Yu$^{1}$, Mar\'ia Mart\'inez Valado$^{1}$, Christoph Hufnagel$^{1}$, Leong Chuan Kwek$^{1,2,3,4}$, Luigi Amico$^{1,5,6}$, and Rainer Dumke$^{1,7}$}\email{rdumke@ntu.edu.sg}


\affiliation{$^{1}$Centre for Quantum Technologies, National University of Singapore, 3 Science Drive 2, Singapore 117543, Singapore}

\affiliation{$^{2}$Institute of Advanced Studies, Nanyang Technological University, 60 Nanyang View, Singapore 639673, Singapore}

\affiliation{$^{3}$National Institute of Education, Nanyang Technological University, 1 Nanyang Walk, Singapore 637616, Singapore}

\affiliation{$^{4}$MajuLab, CNRS-UNS-NUS-NTU International Joint Research Unit, UMI 3654, Singapore}

\affiliation{$^{5}$CNR-MATIS-IMM \& Dipartimento di Fisica e Astronomia, Universit\'a Catania, Via S. Soa 64, 95127 Catania, Italy} 

\affiliation{$^{6}$INFN Laboratori Nazionali del Sud, Via Santa Sofia 62, I-95123 Catania, Italy}

\affiliation{$^{7}$Division of Physics and Applied Physics, Nanyang Technological University, 21 Nanyang Link, Singapore 637371, Singapore}

\begin{abstract}
We investigate a novel hybrid system of a superconducting charge qubit interacting directly with a single neutral atom via electric dipole coupling. Interfacing of the macroscopic superconducting circuit with the microscopic atomic system is accomplished by varying the gate capacitance of the charge qubit. To achieve strong interaction, we employ two Rydberg states with an electric-dipole-allowed transition, which alters the polarizability of the dielectric medium of the gate capacitor. Sweeping the gate voltage with different rates leads to a precise control of hybrid quantum states. Furthermore, we show a possible implementation of a universal two-qubit gate.
\end{abstract}

\pacs{03.67.Lx, 32.80.Qk, 74.50.+r, 85.25.Cp}

\maketitle

\section{Introduction}

Recently, much attention has been paid to hybrid systems composed of superconducting (SC) circuits and neutral atoms~\cite{PRL:Nirrengarten2006,Nature:Colombe2007,PRL:Shimizu2009,PRL:Kubo2010,PRA:Siercke2012,AnnuRevCondensMatterPhys:Daniilidis2013,Nature:Bernon2013,PRA:Qiu2014}. The macroscopic SC devices with the submicrometer-sized Josephson junctions (JJ) possess the advantages of rapid operation ($1\sim10~\mbox{\rm ns}$), scalability, design flexibility, and tunability, but the disadvantage of short relaxation and dephasing times ($0.1\sim10~\mu \mbox{\rm s}$)~\cite{RMP:Makhlin2001,RMP:Xiang2013}. In comparison, the microscopic atomic systems maintain the quantum coherence over long time ($1~\mbox{\rm ms}\sim1~\mbox{\rm s}$) and can be engineered precisely, but they operate slowly ($1\sim10~\mu\mbox{\rm s}$) and have limited scalability~\cite{PRL:Monroe1995,PRL:Brennen1999,PRL:Isenhower2010}. The hybrid systems combined of these two different components have the prospect of rapidly manipulating quantum states and long-time storage of quantum information.

For the transmission of quantum states, the atomic system needs to be entangled with the SC circuits. Linking both components with a quantum bus, i.e., a coplanar waveguide resonator, the SC circuits and atoms are indirectly coupled by the electromagnetic field of the resonator~\cite{PRL:Tian2004,PRL:Rabl2006}. Besides, the atomic system can also directly talk to the SC circuits via the magnetic dipole interaction~\cite{PRL:Marcos2010,RevMexFis:Hoffman2011}. Despite a large number of various proposals for the hybrid systems, the relevant experimental progress is slow. Especially, the direct interface between SC circuits and neutral atoms is still challenging.

For resonant coupling to the solid-state devices, the atomic system should be selected accordingly. In principle, the operating frequency of SC circuits can be matched to the ground-state hyperfine splittings of alkali atoms. In this case the magnetic dipole moment of the atomic system is extremely small, resulting in a weak interface between subsystems. The coupling can be $\sqrt{N}$ enhanced by replacing single atom with an atomic ensemble~\cite{PRL:Patton2013,PRA:Patton2013}, where $N$ is the number of atoms. However, a dense quantum gas causes new issues, such as the atomic number fluctuations and interparticle interactions, making the experimental implementation ambitious.

\begin{figure}[b]
\includegraphics[width=8.5cm]{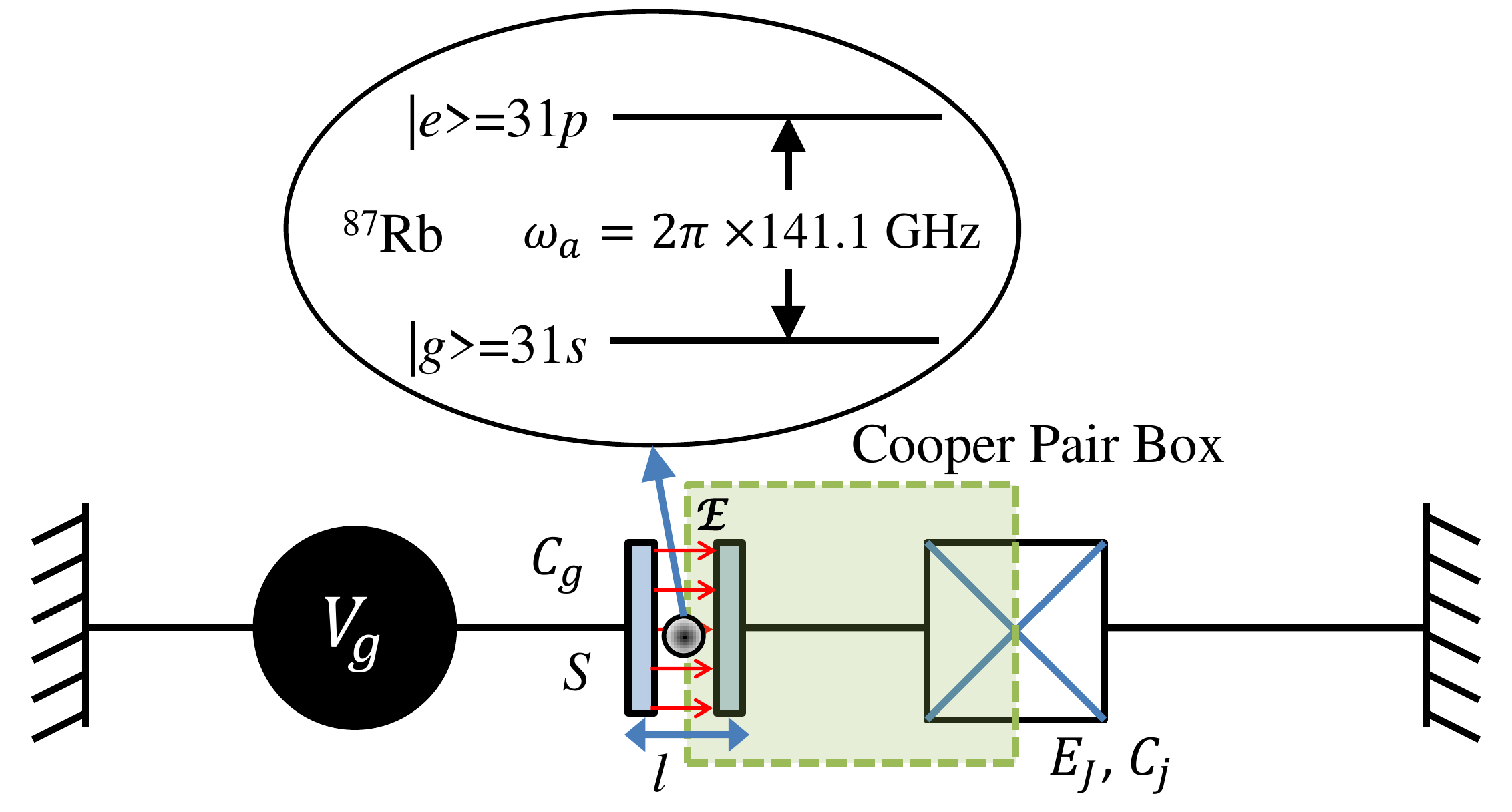}\\
\caption{(Color online) Schematic diagram of a CPB interacting with an atomic qubit. A single $^{87}$Rb atom is placed inside the gate capacitor $C_{g}$ and couples to the electric field $\mathpzc{E}$ between two parallel identical plates with an area $S=4~\mu\mbox{\rm m}\times4~\mu\mbox{\rm m}$ and a separation $l=0.5~\mu\mbox{\rm m}$. The gate capacitance of the empty capacitor (without the extra atom) is $C_{g0}=283.3$ aF and the corresponding empty charging energy is $E_{C0}=2\pi\hbar\times247.4$ GHz. The self-capacitance of the Josephson junction is chosen to be $C_{j}=30$ aF and the Josephson energy is set to $E_{J}=2\pi\hbar\times100$ GHz with the corresponding critical current $I_{c}=200$ nA. The transition frequency between two Rydberg $|g\rangle=31s$ and $|e\rangle=31p$ states is $\omega_{a}=2\pi\times141.1$ GHz and the corresponding electric dipole moment is $d_{eg}=565.8~ea_{0}$.}\label{Fig1}
\end{figure}

In this paper, we study a strongly coupled hybrid system, where a charge qubit directly interacts with an atomic qubit composed of two Rydberg states. Unlike the direct-coupled systems relying on the magnetic dipole interactions~\cite{PRL:Marcos2010,RevMexFis:Hoffman2011,PRL:Patton2013,PRA:Patton2013,PRA:Douce2015}, our system is completely based on the direct electric-dipole interaction between the macroscopic and microscopic quantum systems. This scheme has two advantages: for one thing, the energy separation between two highly excited atomic states can be reduced to hundreds or even tens of GHz, which enables the effective driving by the electromagnetic field at a microwave frequency; for another, the electric-dipole allowed transition between two Rydberg states leads to a large electric dipole moment, which ensures the strong interaction between these two qubits. Our investigation illustrates that the direct electric-dipole coupling in the hybrid system results in a strong interqubit entanglement. The quantum-state control is achieved via linearly sweeping the gate voltage, leading to a universal two-qubit gate.

\section{Physical model}

We consider a Cooper-pair box (CPB) composed of a SC island connected to a SC reservoir via a tunnel junction with a self-capacitance $C_{j}$~\cite{PRL:Nakamura1997,Nature:Nakamura1999,QIP:Pashkin2009} as shown in Fig.~\ref{Fig1}. The Cooper pairs can tunnel from the reservoir to the island through the junction. The SC island is biased by a voltage source $V_{g}$ via a gate capacitor $C_{g}$. For simplicity, we consider $C_{g}$ as a parallel-plate capacitor with the plate area $S$ and the interplate distance $l$. For a common CPB (without coupling to an extra atom), the Hamiltonian including the electrostatic energy, the work done by the voltage source, and the tunneling energy of the Cooper pairs is expressed as
\begin{equation}\label{Hc1}
H_{c}=E_{C}(N-N_{g})^{2}-\frac{C_{g}V^{2}_{g}}{2}-\frac{E_{J}}{2}(e^{i\delta}+e^{-i\delta}),
\end{equation}
with the Coulomb energy of a single Cooper Pair on the island $E_{C}=\frac{(2e)^{2}}{2(C_{g}+C_{j})}$, the number operator of excess Cooper pairs in the box $N$, the offset charge $N_{g}=\frac{C_{g}V_{g}}{2e}$, the Josephson energy $E_{J}=\frac{\Phi_{0}I_{c}}{2\pi}$, and the phase drop across the junction $\delta$. Here $e$ is the electron charge, $\Phi_{0}=\frac{h}{2e}$ is the magnetic flux quantum, $h$ is the Planck constant, and $I_{c}$ is the critical current of the JJ.

In the charge representation, the Hilbert space associated with the SC circuit is spanned by the eigenstates $\{|n\rangle,n=0,1,2,...\}$ of the number operator $N$, $N|n\rangle=n|n\rangle$ and $e^{i\delta}|n\rangle=|n+1\rangle$. At temperatures such that $k_{B}T\ll E_{C}$, we can restrict ourselves to the case of $n\leq1$, for which one further obtains the charge-qubit Hamiltonian~\cite{PhysToday:You2005}
\begin{equation}\label{Hc2}
H_{c}=\frac{E_{C}}{2}(1-2N_{g})\sigma^{(c)}_{z}-\frac{E_{J}}{2}\sigma^{(c)}_{x},
\end{equation}
where the Pauli matrices $\sigma^{(c)}_{z}=|1\rangle\langle1|-|0\rangle\langle0|$ and $\sigma^{(c)}_{x}=|0\rangle\langle1|+|1\rangle\langle0|$. The Hamiltonian~(\ref{Hc2}) indicates that the excess Cooper-pair number $N$ undergoes a Rabi oscillation between 0 and 1 with a frequency of $\Omega_{c}=\hbar^{-1}\sqrt{E^{2}_{C}(1-2N_{g})^{2}+E^{2}_{J}}$ and an amplitude of $N_{c}=E^{2}_{J}/(\hbar\Omega_{c})^{2}$.

The electric field $\mathpzc{E}$ between two parallel plates of the gate capacitor is derived as $\mathpzc{E}=\frac{2e}{(C_{g}+C_{j})l}(N_{j}+N)$ with $N_{j}=\frac{C_{j}V_{g}}{2e}$~\cite{JLowTempPhys:Pekola2012}. It is seen that $\mathpzc{E}$ consists of a constant component proportional to $N_{j}$ and an oscillatory component associated with $N$. The latter can be used to drive another Rabi oscillation of a two-level atom with a transition frequency $\omega_{a}$ nearly equal to $\Omega_{c}$. Moreover, the amplitude $N_{c}$ should be larger than half its maximum height $N_{c}\geq1/2$ to provide a sufficient driving strength. For an effective atom-electric field coupling, the atomic transition frequency $\omega_{a}$ should satisfy $E_{J}\leq\hbar\omega_{a}\leq\sqrt{2}E_{J}$.

We assume a single atom is placed inside the gate capacitor $C_{g}$ and couples to the electric field $\mathpzc{E}$ (see Fig.~\ref{Fig1}). This extra atom, as the dielectric medium, changes the gate capacitance,
\begin{equation}\label{Cg}
C_{g}=(C_{g0}+C_{j})\frac{(N+N_{j})}{N+N_{j}-S\mathpzc{P}/(2e)}-C_{j},
\end{equation}
where the capacitance in the absence of the dielectric is $C_{g0}=\epsilon_{0}S/l$ with the vacuum permittivity $\epsilon_{0}$ and the electric polarization density of the medium $\mathpzc{P}$. The electric field $\mathpzc{E}$ is then given by
\begin{equation}
\mathpzc{E}=\mathpzc{E}_{0}(N+N_{j}-S\mathpzc{P}/(2e)),
\end{equation}
where the electric-field amplitude $\mathpzc{E}_{0}=\frac{2e}{(C_{g0}+C_{j})l}$. As one can see, an extra component associated with the induced electric-dipole moment $\mathpzc{P}$ appears in $\mathpzc{E}$. Moreover, the relevant $C_{g}$-dependent Coulomb energy $E_{C}$ and gate charge bias $N_{g}$ in the Hamiltonian~(\ref{Hc1}) are rewritten as
\begin{eqnarray}
\label{EC}E_{C}&=&E_{C0}\frac{N+N_{j}-S\mathpzc{P}/(2e)}{N+N_{j}},\\
\label{Ng}N_{g}&=&(N_{g0}+N_{j})\frac{N+N_{j}}{N+N_{j}-S\mathpzc{P}/(2e)}-N_{j},
\end{eqnarray}
where the charging energy constant $E_{C0}=\frac{(2e)^{2}}{2(C_{g0}+C_{j})}$ and the charge bias $N_{g0}=\frac{C_{g0}V_{g}}{2e}$ corresponding to the empty gate capacitance $C_{g0}$.

We consider the single-atom dielectric composed of two atomic states $|g\rangle$ (lower) and $|e\rangle$ (upper) with the electric dipole-allowed transition frequency $\omega_{a}$. Using the $x$-component Pauli matrix $\sigma^{(a)}_{x}=|e\rangle\langle g|+|g\rangle\langle e|$, the atomic polarization operator is written as $\mathpzc{P}=\frac{d_{eg}}{Sl}\sigma^{(a)}_{x}$, where $d_{eg}$ is the electric dipole moment and the direction of the electric field $\mathpzc{E}$ is chosen as the quantization axis. The Hamiltonian for the single atom is expressed as
\begin{equation}
H_{a}=\frac{\hbar\omega_{a}}{2}\sigma^{(a)}_{z}+\hbar\Omega_{a}(N+N_{j})(1+\sigma^{(a)}_{x}),
\end{equation}
with the $z$-component Pauli matrix $\sigma^{(a)}_{z}=|e\rangle\langle e|-|g\rangle\langle g|$ and the Rabi frequency $\Omega_{a}=-\frac{d_{eg}\mathpzc{E}_{0}}{\hbar}$.

The Hamiltonian for the whole hybrid system is given by $H=H_{c}+H_{a}$ with the corresponding Hilbert space spanned by the orthonormal basis $\{|u,n\rangle;u=e,g;n=0,1,2,...\}$. In the limit of $k_{B}T\ll E_{C0}$, we arrive at a direct-coupled system composed of a charge qubit and an atomic qubit. Diagonalizing $H$,
\begin{equation}
H\Psi_{k}=(\mathcal{E}_{k}+C_{g0}V^{2}_{g}/2)\Psi_{k},
\end{equation}
gives the energy spectrum $\mathcal{E}_{k}$ and eigenstates $\Psi_{k}$, where one needs to use the following relation
\begin{equation}
\frac{N+N_{j}}{N+N_{j}-S\mathpzc{P}/(2e)}=\frac{(N+N_{j})^{2}}{(N+N_{j})^{2}-N^{2}_{d}}\left(1+\frac{N_{d}}{N+N_{j}}\sigma^{(a)}_{x}\right),
\end{equation}
with $N_{d}=\frac{d_{eg}}{(2e)l}$.

\begin{figure}
\includegraphics[width=7.5cm]{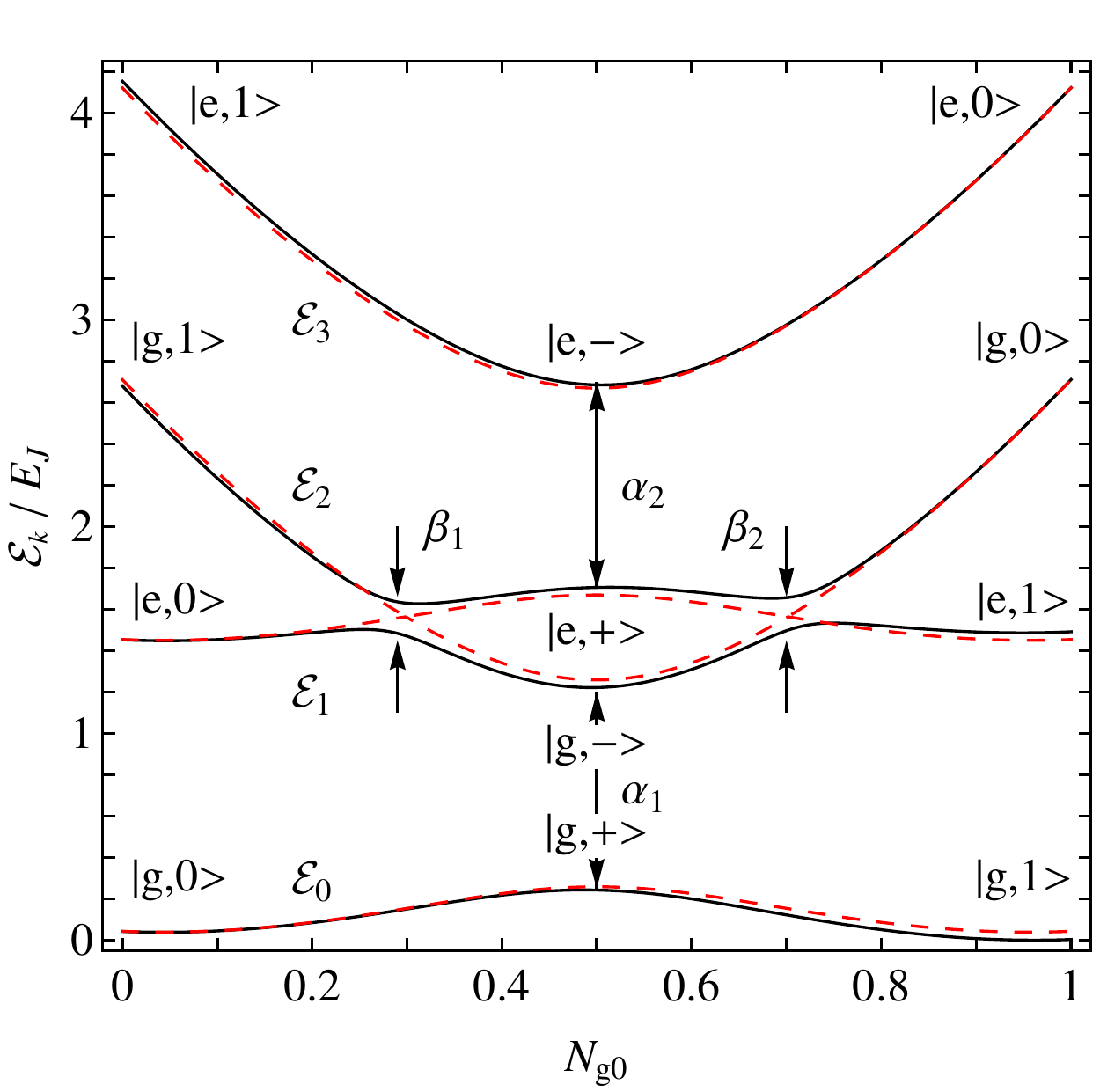}\\
\caption{(Color online) Eigenenergy bands $\mathcal{E}_{k}$ vs. $N_{g0}$. The solid curves depict the numerical results from diagonalizing $H$. The dashed lines give the asymptotic behavior ($d_{eg}=0$). $\alpha_{1,2}$ and $\beta_{1,2}$ mark two anti-crossings caused by the Josephson tunneling at $N_{g0}=0.5$ and two anti-crossings induced by the interqubit coupling at $N_{g0}\approx0.3$ and $0.7$, respectively. The system eigenstates at $N_{g0}=0$, 0.5, and 1 are labeled, where $|u,\pm\rangle=\frac{1}{\sqrt{2}}\left(|u,0\rangle\pm|u,1\rangle\right)$ with $u=e,g$.}\label{Fig2}
\end{figure}


We assume that the area of the gate-capacitor plates is $S=4~\mu\mbox{\rm m}\times4~\mu\mbox{\rm m}$ and the plate distance is $l=0.5~\mu\mbox{\rm m}$, which yields the empty gate capacitance $C_{g0}=283.3$ aF. The small self-capacitance of the JJ is chosen to be $C_{j}=30$ aF, resulting in the Coulomb-energy constant $E_{C0}=2\pi\hbar\times247.4$ GHz and the electric-field amplitude $\mathpzc{E}_{0}=20.5$ V/cm. The Josephson energy is set at $E_{J}=2\pi\hbar\times100$ GHz with the critical current $I_{c}=200$ nA.

The single $^{87}$Rb atom is employed as the atomic qubit placed inside the gate capacitor. We focus on the electric dipole-allowed Rydberg-Rydberg transition between $|g\rangle=31s$ and $|e\rangle=31p$ states with the transition frequency $\omega_{a}=2\pi\times141.1$ GHz~\cite{PRA:Li2003}. The corresponding electric dipole moment is given by $d_{eg}=|\langle31s||d||31p\rangle|=565.8~ea_{0}$~\cite{PRA:Zimmerman1979,JPB:Ryabtsev2005}, where $a_{0}$ is the Bohr radius, leading to the Rabi frequency $\Omega_{a}=2\pi\times14.8$ GHz.

We should note that the effect of other atomic transitions associated with $31s$ and $31p$ can be neglected due to the off-resonant one- and two-photon transitions and the weak-field couplings. In addition, the fine-structure interaction of $31p$, which is of the order of a few GHz~\cite{PRA:Li2003}, is much smaller compared with the strong CPB-atom coupling. Moreover, in a cryogenic environment the radiative lifetimes of $31s$ and $31p$ are 25.3 $\mu\mbox{\rm s}$ and 67.2 $\mu\mbox{\rm s}$~\cite{JPB:Branden2010}, respectively, which are long enough for coherently interfacing the CPB and the atom. The coupling is in principle switchable by coherently transferring two Rydberg states to the hyperfine ground states via the standard spectroscopic techniques, extending the coherence time of atomic qubit~\cite{PRA:Pritchard2014}. The investigated coherent dynamics of hybrid system is within a time scale smaller than the decoherence time of charge qubit (see below).

\section{Quantum computing}

We show in Fig.~\ref{Fig2} the dependence of the eigenenergies $\mathcal{E}_{k}$ on the offset charge $N_{g0}$ which is tuned from 0 to 1 by the gate voltage $V_{g}$. It is seen that the whole energy structure consists of two sets of the charge-qubit-like spectra with the Josephson-tunneling-induced anti-crossings labeled as $\alpha_{1,2}$, whose energy spacings are approximately equal to $\hbar\Delta_{1}=E_{J}$ at $N_{g0}=0.5$. Besides, two other avoided crossings labeled as $\beta_{1,2}$ occur between $\mathcal{E}_{1}$ and $\mathcal{E}_{2}$. We find that $\beta_{1,2}$ are located at $N_{g0}\approx0.3$ and $N_{g0}\approx0.7$, respectively, and the energy spacings are approximately equal to $\hbar\Delta_{2}\approx0.22E_{J}$.

Moreover, we have labeled the eigenstates $\Psi_{k}$ at several specific $N_{g0}$ in Fig.~\ref{Fig2}. At $N_{g0}=0$, the hybrid system can be described by $\Psi_{0}\approx|g,0\rangle$, $\Psi_{1}\approx|e,0\rangle$, $\Psi_{2}\approx|g,1\rangle$, and $\Psi_{3}\approx|e,1\rangle$ for different energy bands. By contrast, for $N_{g0}=1$ the eigenstates are given by $\Psi_{0}\approx|g,1\rangle$, $\Psi_{1}\approx|e,1\rangle$, $\Psi_{2}\approx|g,0\rangle$, and $\Psi_{3}\approx|e,0\rangle$. In particular, at $N_{g0}\simeq0.5$, we find that the system is almost in the charge superposition states of $\Psi_{0,1}\approx|g,\pm\rangle$ and $\Psi_{2,3}\approx|e,\pm\rangle$.

\begin{figure*}
\includegraphics[width=4.25cm]{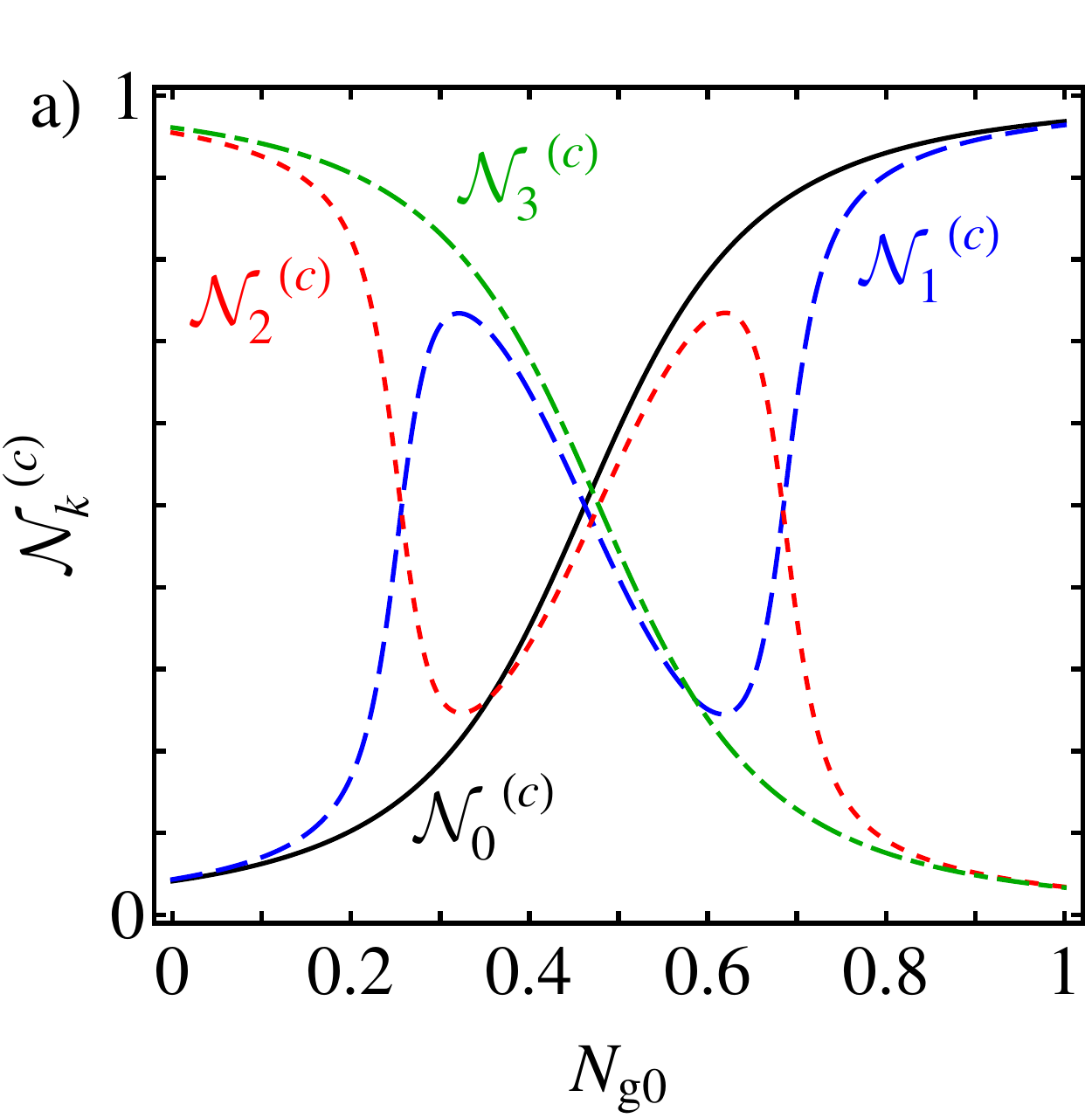}~\includegraphics[width=4.3cm]{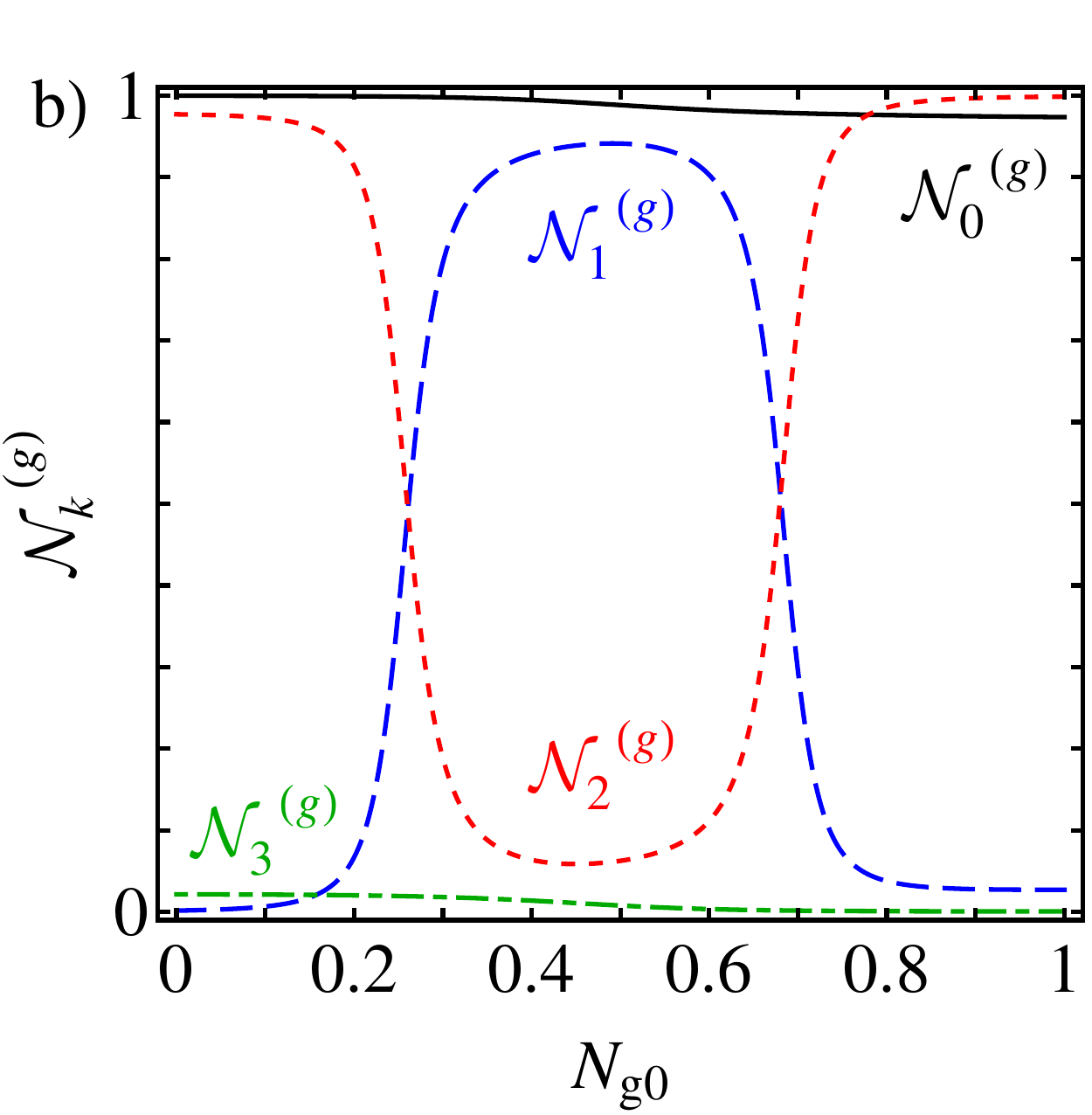}~\includegraphics[width=4.3cm]{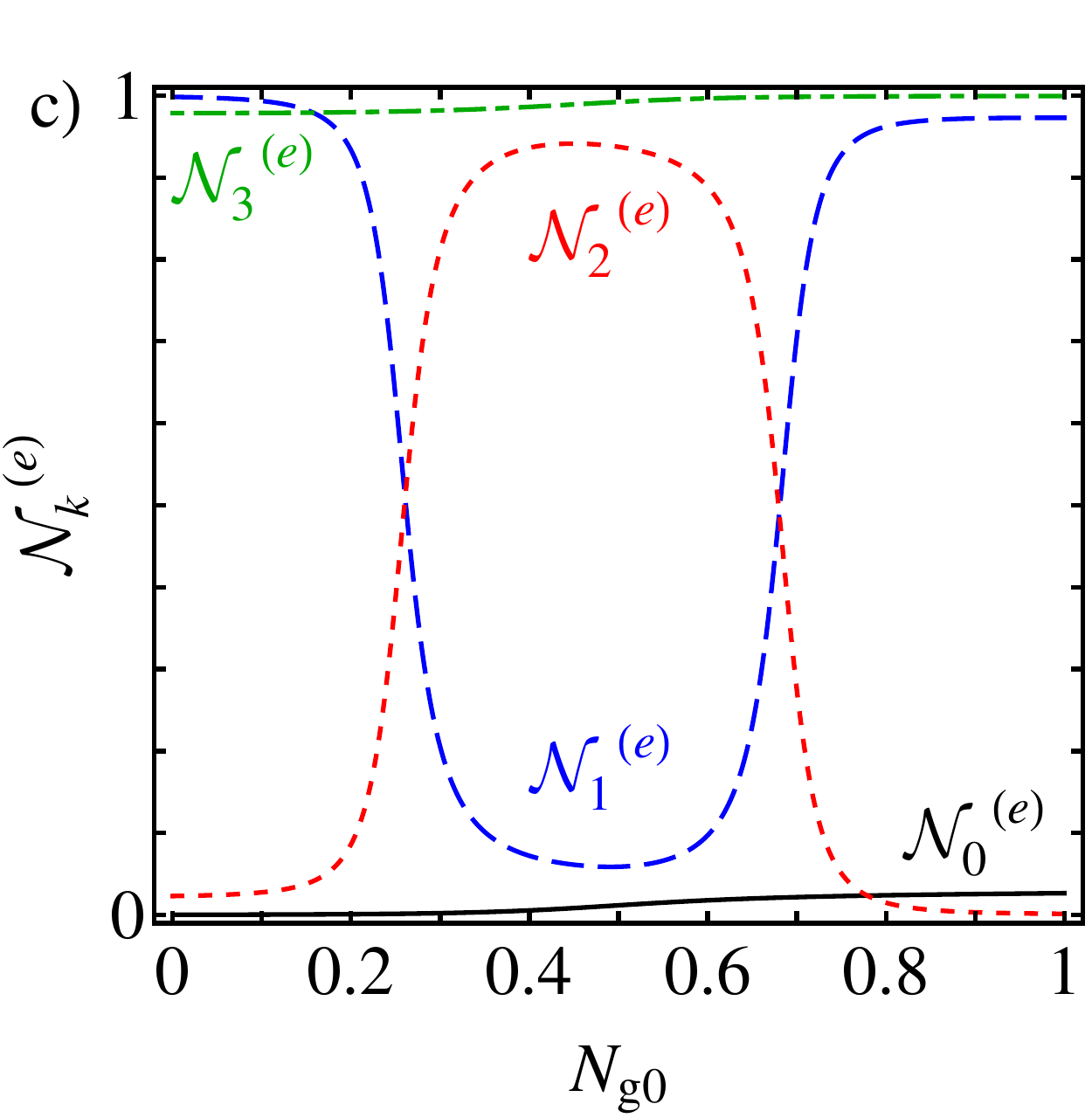}~\includegraphics[width=4.3cm]{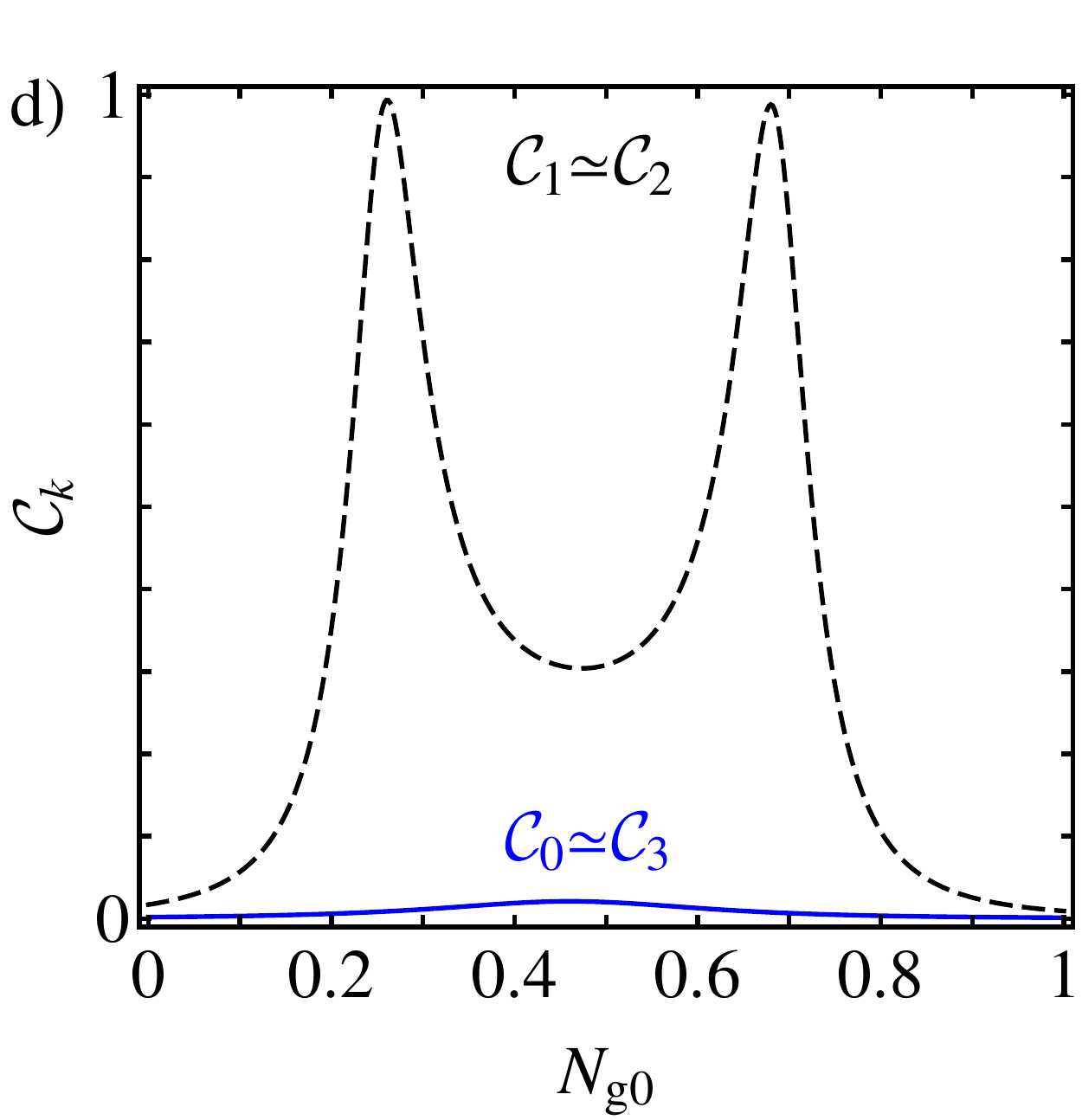}\\
\caption{(Color online) The expectation values $\mathcal{N}^{(c)}_{k}$ of the excess number of Cooper pairs in the box and the atomic populations $\mathcal{N}^{(u)}_{k}$ in $|u=g,e\rangle$ are plotted in (a), (b), and (c), respectively. (d) The concurrences $\mathcal{C}_{k}$ for the two-qubit system in different $\Psi_{k}$. In the bands of $k=1$ and $k=2$ the charge-atom entanglement maximizes at two anticrossings $\beta_{1,2}$ while two qubits are weakly coupled with each other in the bands of $k=0$ and $k=3$.}\label{Fig3}
\end{figure*}

\subsection{Qubit entanglement}

From $\Psi_{k}$, one can also derive the expectation values of the excess number of Cooper pairs in the box, $\mathcal{N}^{(c)}_{k}=\langle\psi_{k}|N|\psi_{k}\rangle$, as shown in Fig.~\ref{Fig3}a. We see that $\mathcal{N}^{(c)}_{0}$ ($\mathcal{N}^{(c)}_{3}$) behaves like a usual charge qubit, i.e., monotonically increasing (decreasing) from 0 (1) to 1 (0)~\cite{Book:Vion2004}, while $\mathcal{N}^{(c)}_{1,2}$ exhibit the opposed oscillatory behavior. In Fig.~\ref{Fig3}b and c, we display the average atomic populations in $|g\rangle$ and $|e\rangle$, $\mathcal{N}^{(u)}_{k}=\langle\Psi_{k}|u\rangle\langle u|\Psi_{k}\rangle$. Passing through either avoided crossing $\beta_{1,2}$, the atom flips between $|g\rangle$ and $|e\rangle$ in the bands of $k=1$ and $k=2$ while the corresponding Cooper-pair number varies abruptly, which indicates the energy exchange between the charge and atomic qubits and strong interqubit entanglement. By contrast, in the bands of $k=0$ and $k=3$ the atom mainly stays in $|g\rangle$ and $|e\rangle$, respectively, meaning weak entanglement between two qubits. We use the concurrence $\mathcal{C}_{k}$ to measure the entanglement~\cite{PRL:Hill1997,PRL:Wootters1998} of the two-qubit system in different bands (see Fig.~\ref{Fig3}d). As expected for the bands $k=0$ and $k=3$, $\mathcal{C}_{0,3}\sim0$ indicate that the atom weakly couples to the charge qubit and $\Psi_{0,3}$ can be approximately written as product states. Contrarily, the charge and atomic qubits are strongly entangled in $\Psi_{1,2}$ corresponding to the bands $k=1$ and $k=2$. Around $\beta_{1,2}$ we obtain $\mathcal{C}_{1,2}\sim1$, i.e., the maximal interqubit entanglement, which corresponds to the strong charge-number variation (Fig.~\ref{Fig3}a) and the atomic population-flipping (Fig.~\ref{Fig3}b and c).

\subsection{Qubit control}

To enable transitions between different qubit states, we sweep the hybrid system through the avoided crossings, similar to the Landau-Zener-St\"uckelberg (LZS) interferometry~\cite{Science:Oliver2005,PRB:Sun2011,PRB:Stehlik2012,SciRep:Sun2015}. As an example, we numerically perform a single-passage sweep with the system initially prepared in $\psi_{i}=|e,0\rangle$ by linearly varying the offset charge $N_{g0}$ (via changing the gate voltage $V_{g}$) from 0 to 1 with a constant rate $v$, and check the probabilities $P_{u,n}$ of finding the system in different $|u,n\rangle$.

For our hybrid system, sweeping the system through the avoided crossings $\beta_{1,2}$ is related to the LZS interferometry in a single SC qubit, where the quantum system is swept through the same single anti-crossing twice~\cite{Science:Oliver2005}. Analogous to an optical Mach-Zehnder interferometer, two anticrossings $\beta_{1,2}$ act like two beamsplitters. At $\beta_{1}$, the entrance state is split into two, which travel along two paths $\mathcal{E}_{1}$ and $\mathcal{E}_{2}$, respectively (see Fig.~\ref{Fig2}). Afterwards, those two states are combined together at $\beta_{2}$ and interfere with each other. Nevertheless, we should point out that our coupled system is significantly different from the usual LZS interferometry performed on a single SC qubit. Due to the avoided crossings $\alpha_{1,2}$, the adiabatic states of two paths $\mathcal{E}_{1}$ and $\mathcal{E}_{2}$ are not restricted to two states $|e,0\rangle$ and $|g,1\rangle$. Thus, the conventional analysis based on the two-state Landau-Zener model is not suitable.

\begin{figure*}
\includegraphics[width=4.3cm]{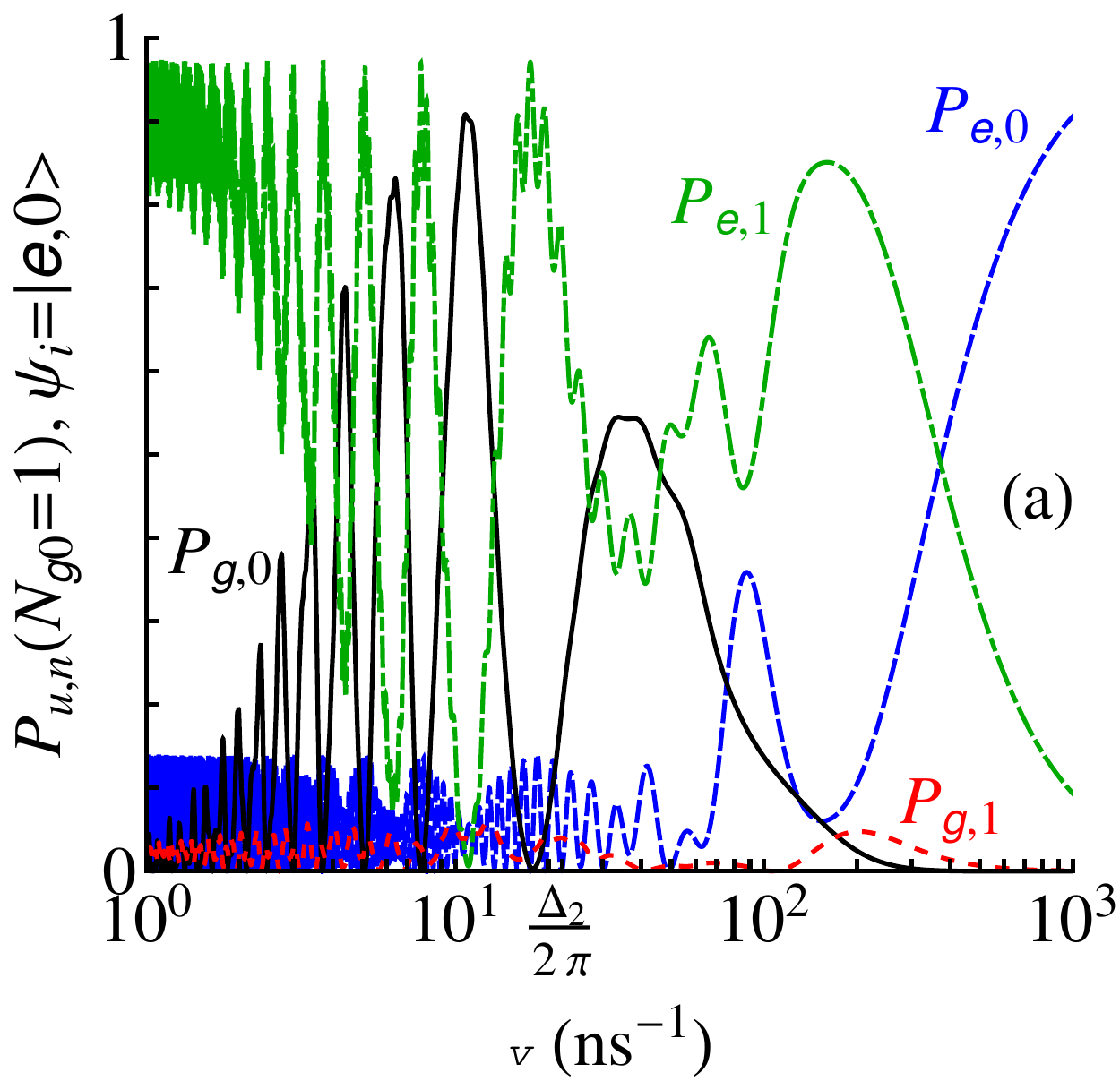}~\includegraphics[width=4.3cm]{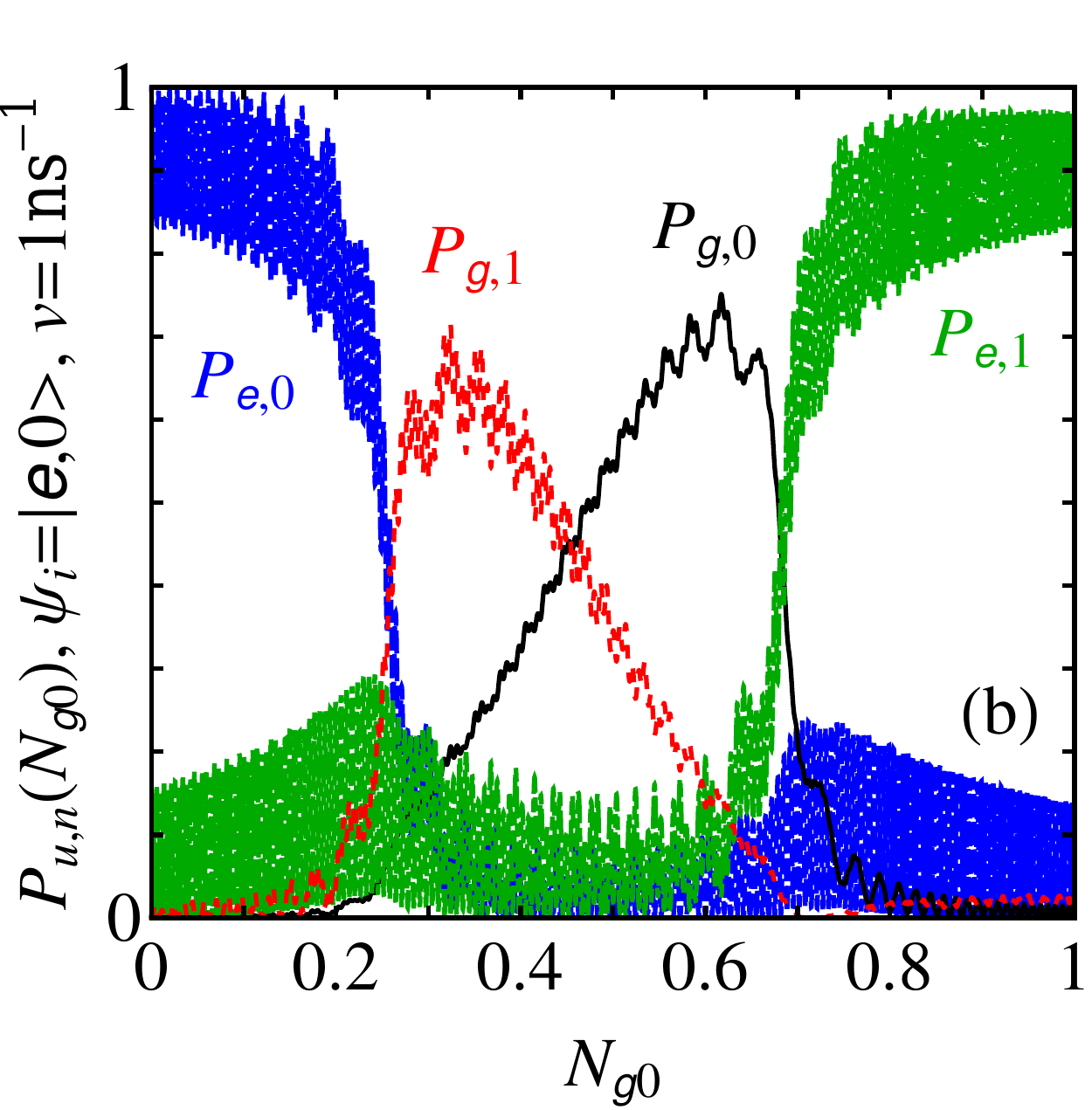}~\includegraphics[width=4.3cm]{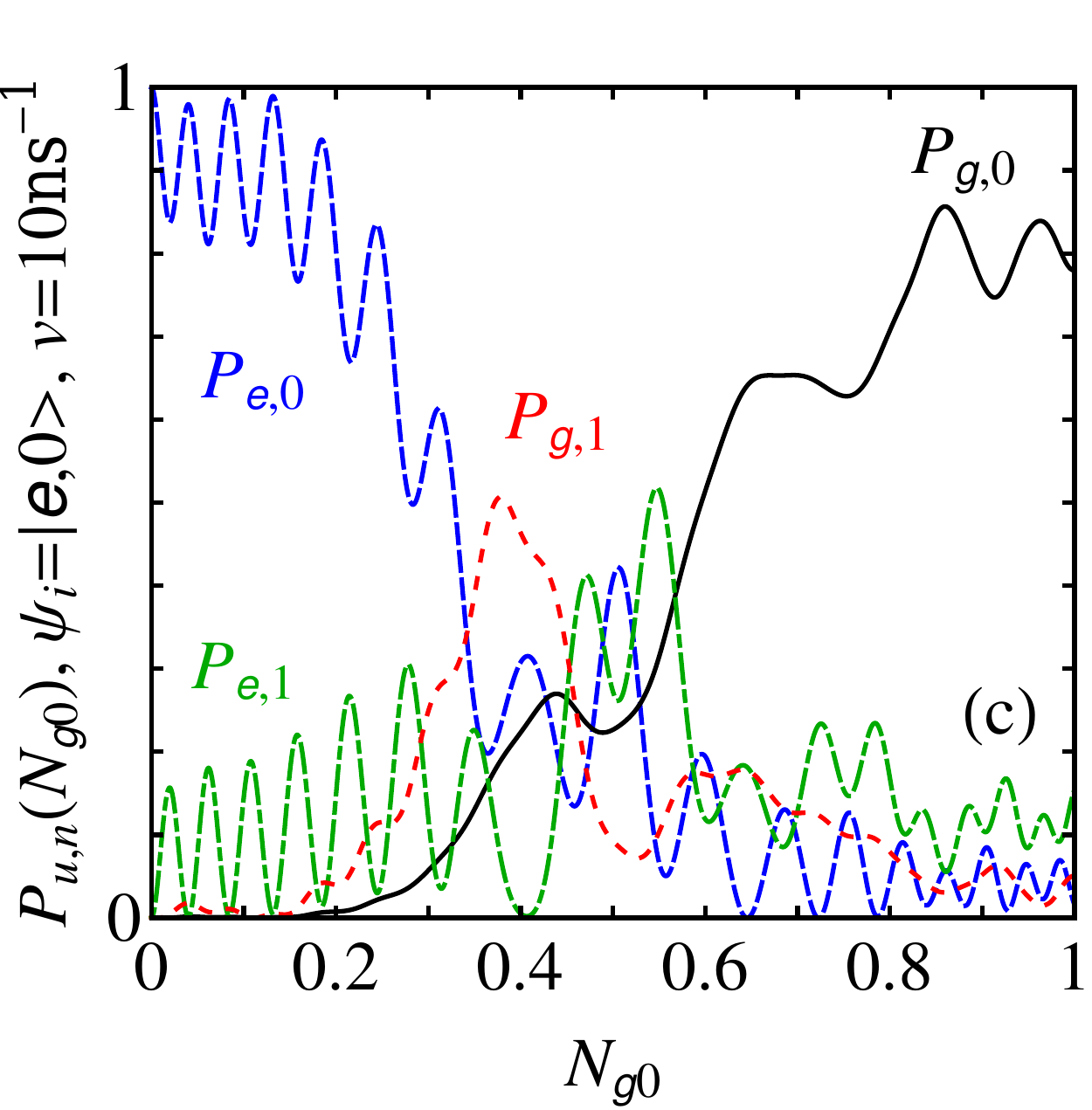}~\includegraphics[width=4.3cm]{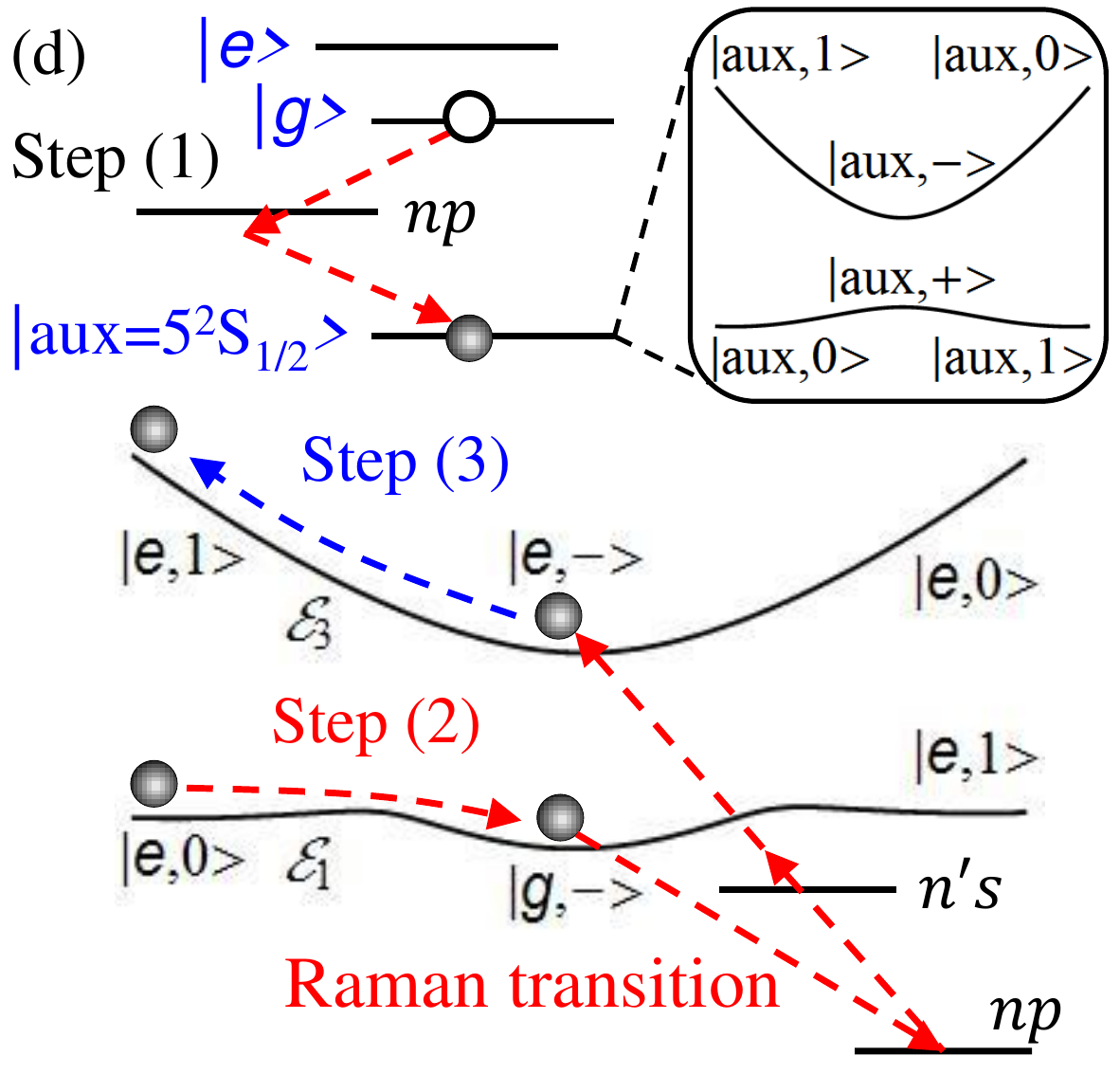}\\
\caption{(Color online) Single-passage sweep of the hybrid system with an initial state $\psi_{i}=|e,0\rangle$. (a) The probabilities $P_{u,n}$ of the system at $N_{g0}=1$ in different basis states $|u,n\rangle$ as a function of the sweep rate $v$. Three regimes of the sweep rate $v$ can be identified: In the limit of $v\ll\Delta_{2}/(2\pi)$, the system at $N_{g0}=1$ mainly remains in $|e,1\rangle$. In the intermediate regime $v\sim\Delta_{2}/(2\pi)$, $P_{g,0}$ and $P_{e,1}$ at $N_{g0}=1$ oscillate strongly. For $v\gg\Delta_{2}/(2\pi)$, the system is transferred to $|e,0\rangle$. Two examples of $P_{u,n}$ vs. $N_{g0}$ with $v=1$ ns$^{-1}$ and $v=10$ ns$^{-1}$ are shown in (b) and (c), respectively. (d) Three steps of the CNOT gate. The atom in $|g\rangle$ is transferred in an auxiliary state, for example $|\textrm{aux}\rangle=5^{2}S_{1/2}$, in step 1. In step 2, the gate charge bias is swept adiabatically from 0 to 0.5 and the transition between bands $k=1$ and $k=3$ is accomplished via the Raman transition. The charge bias is swept back to 0 and the atom in $|\textrm{aux}\rangle$ is brought back to $|g\rangle$ in step 3.}\label{Fig4}
\end{figure*}

We numerically solve the time-dependent Schr\"odinger equation and plot the results in Fig.~\ref{Fig4}a. It is seen that the final state of the system strongly depends on the sweep rate $v$. In the limit of $v\ll\Delta_{2}/(2\pi)$, the system at $N_{g0}=1$ is mainly in $|e,1\rangle$, which indicates that the coupled system follows the path $\mathcal{E}_{1}$ adiabatically. As illustrated in Fig.~\ref{Fig4}b with a sweep rate $v=1$ ns$^{-1}$, at the anti-crossing $\beta_{1}$, the system has a high probability in $|g,1\rangle$ and then transits to the superposition state $|g,-\rangle$ at $N_{g0}=0.5$ (see Fig.~\ref{Fig2}). Finally, the system is transfered to $|e,1\rangle$ at the anticrossing $\beta_{2}$.

As $v$ approaches the energy separation $\Delta_{2}/(2\pi)$, the system state at $N_{g0}=1$ displays an oscillation between $|e,1\rangle$ and $|g,0\rangle$ (see Fig.~\ref{Fig4}a), which results from the interference between two paths of $\mathcal{E}_{1}$ and $\mathcal{E}_{2}$. As shown in Fig.~\ref{Fig4}c with a sweep rate $v=10$ ns$^{-1}$, at $\beta_{1,2}$ the state transformation of the system is not distinct. Around $N_{g0}=0.5$ four states $|u,n\rangle$ with $u=e,g$ and $n=\pm1$ are mixed together, and finally $|g,0\rangle$ gets constructively enhanced. When $v$ moves away from $\Delta_{2}/(2\pi)$, this interference effect is weakened. The system can non-adiabatically pass though the anti-crossings $\beta_{1,2}$. Finally, for $v>\Delta_{1}/(2\pi)=10^{2}$ ns$^{-1}$, the system is mainly in $|e,0\rangle$, indicating the system non-adiabatically passes though the avoided crossing $\alpha_{1}$.

\subsection{Two-qubit logic gates}

Naturally, this hybrid system is also applicable for quantum computation~\cite{Nature:Bennett2000}. As an illustration, we consider the implementation of the controlled-NOT (CNOT) gate~\cite{PRL:Isenhower2010}, where the atomic and charge qubits play the roles of control and target, respectively. As shown in Fig.~\ref{Fig4}d, the concrete operation is demonstrated according to the following steps: (1) Initially, the offset charge $N_{g0}$ is set to be zero. Two light pulses are applied on the two-photon transition between $|g\rangle$ and an auxiliary state $|\textrm{aux}\rangle$, for instance, $5^{2}S_{1/2}$, so as to completely transfer the occupation in $|g\rangle$ (if any) to $|\textrm{aux}\rangle$~\cite{PRL:Monroe1995}. (2) $N_{g0}$ is swept adiabatically ($v\ll\Delta_{2}/(2\pi)$) from 0 to 0.5, and a Raman transition between $|g,-\rangle$ and $|e,-\rangle$ is implemented in rapid succession. (3) $N_{g0}$ is swept adiabatically back to zero and afterwards the two-photon transition between $|g\rangle$ and $|\textrm{aux}\rangle$ is applied again to transfer the atom in $|\textrm{aux}\rangle$ back to $|g\rangle$.

For the system initially prepared in $|e,0\rangle$ ($|e,1\rangle$) at $N_{g0}=0$, the step (1) does not affect the state (see Fig.~\ref{Fig4}d). In step (2), the system adiabatically evolves to $|g,-\rangle$ ($|e,-\rangle$) and then transits to $|e,-\rangle$ ($|g,-\rangle$) via the Raman process. Finally, the system adiabatically moves to $|e,1\rangle$ ($|e,0\rangle$) after step (3). By contrast, for the system initially prepared in $|g,0\rangle$ or $|g,1\rangle$, step (1) brings the system in a space spanned by $|\textrm{aux},0\rangle$ and $|\textrm{aux},1\rangle$, which are uncoupled to any $|u,n\rangle$ due to the optical-frequency separation. In addition, the Raman transition in step (2) does not take effect, and the system still stay in its original state after the whole operation. Therefore, the CNOT operation is executed.

\section{Coherence property}

For our strongly-coupled system, the finite lifetimes of two atomic states and the relaxation and dephasing of superconducting circuit limit the coherent interaction between two different quantum subsystems.

Although the Rydberg states are characterized by their extremely large electric dipole moments and relatively long lifetimes (typically, tens of $\mu$s), they are not ideal candidates for storing quantum information. The Rydberg atom can work as an intermediate qubit to achieve the strong inter-qubit entanglement and implement the two-qubit logic gates within a time scale much shorter than the Rydberg-state lifetimes. After the quantum control operations, one can map two Rydberg states onto two hyperfine ground states which are usually employed for long-time storage of quantum states~\cite{PRA:Pritchard2014}.

Due to the strong coupling to the environment, the superconducting circuits suffer from the short relaxation time $T_{1}$ and dephasing time $T_{2}<T_{1}$. For the common charge qubit, the dominant $1/f$ noise in background charge damps the quantum coherent dynamics of cooper pairs on the island after about 10 ns~\cite{NewJPhy:Bladh,Science:Devoret}. Nevertheless, as long as the gate operation times are shorter than the decoherence time of superconducting circuit, the relaxation and dephasing of charge qubit hardly affect the system dynamics. For our strongly-coupled system, the qubit control can be performed well within 1 ns (Fig.~\ref{Fig4}a). Moreover, the two-qubit logic gates (Fig.\ref{Fig4}d) can be also implemented in a similar time scale.

\section{Conclusion}

In summary, we have investigated a hybrid system composed of a charge qubit and an atomic qubit. The oscillation of the excess Cooper pairs in the box varies the internal electric field of the gate capacitor and further directly drives the rotation of the atomic qubit. The strong interqubit coupling is achieved by choosing the atomic electric-dipole transition between two Rydberg states with a frequency spacing that is nearly resonant with the operating frequency of the SC circuit. As a result, we obtain strong entanglement between two different kinds of qubits. The quantum states of this two-qubit system can be controlled by sweeping the gate voltage with different rates. Moreover, universal two-qubit logic gates can be implemented, showing the potential for quantum state transfer and computation.

In this work, we have assumed an ideal CPB in the absence of decoherence sources, which is valid within a time scale shorter than the decoherence times of qubits. In reality, due to the strong coupling to the local electromagnetic environment, the quantum dynamics of the charge qubit is strongly influenced by the $1/f$ noise in background charge and critical current~\cite{PRL:Aassime2001} and even the readout back-action~\cite{PRB:Schreier2008}, resulting in a short decoherence time. In the future study, we will investigate how the external noise affects the implementation of quantum state control on this hybrid system.

\emph{Acknowledgment:} This research is supported by the National Research Foundation Singapore under its Competitive Research Programme (CRP Award No. NRF-CRP12-2013-03) and the Centre for Quantum Technologies, Singapore.

\end{document}